\documentclass[11pt]{article}
\usepackage[letterpaper,margin=0.72in]{geometry}
\usepackage{amsmath,amssymb,bm}
\usepackage{graphicx}
\usepackage{booktabs,tabularx,array,longtable}
\usepackage{microtype}
\usepackage{authblk}
\usepackage[hidelinks]{hyperref}
\usepackage{caption}
\usepackage{enumitem}
\usepackage{setspace}
\usepackage{float}
\newcommand{\erg}{V_{\mathrm{ERG}}}

\newcommand{\pc}{p_c^{(200)}}

\title{\textbf{Causal Multiscale Modeling of Ocular FLASH Electroretinography: Model Development, Experimental Fitting, and Spin-Glass--Percolation Predictions}}
\author[1,2,*]{Ramin Abolfath}
\author[3]{Mohammad Rezaee}
\affil[1]{Department of Physics and Astronomy, Howard University, Washington, DC 20059, USA}
\affil[2]{Department of Radiation Physics and Oncology, University of Texas MD Anderson Cancer Center, Houston, TX 75031, USA}
\affil[3]{Department of Radiation Oncology and Molecular Radiation Sciences, Johns Hopkins University School of Medicine, Baltimore, Maryland, USA}
\affil[*]{Corresponding author: Ramin Abolfath}
\date{}
\begin{document}
\maketitle

\begin{abstract}
\textbf{Purpose:} To develop a causal multiscale model of retinal electroretinography (ERG) after ocular ultrahigh-dose-rate (FLASH) and conventional-dose-rate (CONV) irradiation, evaluate the framework against published ocular ERG data, and derive prospective dose-response predictions for future validation. \textbf{Methods:} The retina was coarse-grained into $N=200$ functional modules. Local rate equations were formulated as Hodgkin--Huxley-inspired conductance-balance/RC reductions of retinal ion-channel and synaptic pathways. Radiation generated an injury field that altered phototransduction, a-wave and b-wave dynamics and, in the network extension, inter-module coupling and bond survival. A reduced amplitude benchmark and nested M0/M1/M2 causal models were evaluated. \textbf{Results:} The reduced benchmark achieved $R^2=0.9721$ and RMSE $=11.94~\mu$V. M0, M1, and M2 yielded $R^2=0.8546$, 0.8559, and 0.8560, respectively, indicating limited independent identifiability of the network terms from amplitude data alone. Under an exploratory 50\% retained-gain criterion, the reduced benchmark predicted a-wave preservation thresholds of 28.6 Gy for FLASH and 23.7 Gy for CONV, and b-wave thresholds of 29.2 Gy and 24.1 Gy, respectively, corresponding to a model-based functional dose-modifying factor (DMF) of 1.21. \textbf{Conclusion:} The ERG amplitude data support the causal radiation-injury-to-ERG framework, while additional observables will be needed to distinguish the predicted collective-network contributions. The model also provides quantitative dose targets for prospective studies of the FLASH retinal-preservation window and a framework for testing how local retinal injury may propagate into collective dysfunction.
\end{abstract}

\section{Introduction}
Radiation therapy is a major eye-preserving treatment for localized uveal melanoma. In the Collaborative Ocular Melanoma Study (COMS), $^{125}$I plaque brachytherapy for medium-sized choroidal melanoma produced 5-year survival comparable with enucleation (82\% vs 81\%) \cite{COMS2001}. American Brachytherapy Society guidance recommends a minimum tumor dose of 85 Gy for $^{125}$I plaque therapy in appropriate patients \cite{Nag2003}, while proton-beam therapy is commonly delivered with highly hypofractionated schedules such as 60 cobalt-gray equivalent in four fractions \cite{Desjardins2012}. Normal-tissue toxicity nevertheless limits visual preservation; nonproliferative radiation retinopathy was estimated in 42\% of patients by 5 years in a large plaque-radiotherapy series \cite{Gunduz1999}. These observations motivate models that connect physical radiation delivery to retinal function.

Ultrahigh-dose-rate FLASH radiotherapy may widen this normal-tissue therapeutic window, although ocular tumor control under FLASH remains to be established. In the preclinical ocular study of Miles et al., healthy mouse eyes received 21 or 26 Gy under FLASH or conventional-dose-rate (CONV) irradiation, followed by ERG and histopathology at two months \cite{Miles2025}. At 26 Gy, retinal function was preserved after FLASH but impaired after CONV irradiation, with corresponding structural injury after CONV \cite{Miles2025}. The X-ray FLASH platform and dosimetry were characterized previously \cite{Rezaee2021}. Ocular FLASH therefore provides a useful setting in which radiation delivery, retinal morphology, and electrophysiologic function can be linked within one organ.

ERG provides a natural multiscale readout: the negative a-wave is dominated by photoreceptor activity, whereas the positive b-wave reflects mainly post-photoreceptor signaling, particularly ON-bipolar-cell activity \cite{Robson2022,Frishman2006,Stockton1989}. At the membrane level, these responses arise from ion-channel currents and synaptic conductances. The present rate equations use a Hodgkin--Huxley conductance-balance foundation \cite{Hodgkin1952}, reduced to effective first-order RC dynamics appropriate for graded retinal responses. The coarse-grained pathways include the rhodopsin--transducin--PDE6--cGMP cascade and CNG Na$^+$/Ca$^{2+}$ current, photoreceptor K$^+$ currents, Ca$_V$1.4-dependent ribbon-synaptic glutamate release, and the ON-bipolar mGluR6--G$_o$--TRPM1 pathway \cite{FuYau2007,Heidelberger2005,Masu1995,Morgans2009,Martemyanov2017,VanHook2019}.

In this work we report developing a causal $N=200$ coarse-grained model in which $\erg(0)=0$ and a- and b-wave timing emerges dynamically. Radiation injury modifies local gains and time constants and, at the network level, disordered inter-module couplings and functional bond survival. These collective components use concepts from spin-glass theory \cite{Edwards1975}, stochastic Ising dynamics \cite{Glauber1963}, random graphs \cite{Gilbert1959,Erdos1959}, and finite-size percolation \cite{Stauffer1994}; each Ising variable represents a coarse-grained retinal population, not a single neuron or literal magnetic spin.

The present work combines two fitting approaches. A reduced amplitude-level model provides a compact benchmark for stimulus-response and radiation-injury dependence, whereas nested M0/M1/M2 realizations of the complete $N=200$ model test whether disordered coupling and percolative bond loss are distinguishable from local injury. We also use the reduced benchmark to derive a prospective functional-preservation dose threshold and ocular dose-modifying factor (DMF). This calculation is intended to generate a falsifiable dose-escalation hypothesis, not a clinical tolerance dose.

\section{Methods}
\subsection{Experimental constraints}
ERG intensity-response curves from the published ocular FLASH study were used as the experimental constraints \cite{Miles2025}. Briefly, healthy female C57BL/6J mouse eyes were irradiated with single fractions of 21 or 26 Gy. The measured dose rates were $52.6\pm2.6$ Gy/s for FLASH and $1.1\pm0.1$ Gy/s for CONV; retinal function was assessed two months after irradiation using ERG, with histopathology providing an independent structural endpoint \cite{Miles2025}. The published ERG analysis reported five eyes per group. In the present work, fits are performed to the reported group-mean amplitude-versus-intensity responses. Consequently, the fitted parameters are interpreted primarily as tests of model architecture and as quantitative predictions for prospective validation rather than as definitive eye-level biological constants.

\subsection{Causal multiscale architecture}
\begin{figure}[t]
\centering
\includegraphics[width=\linewidth]{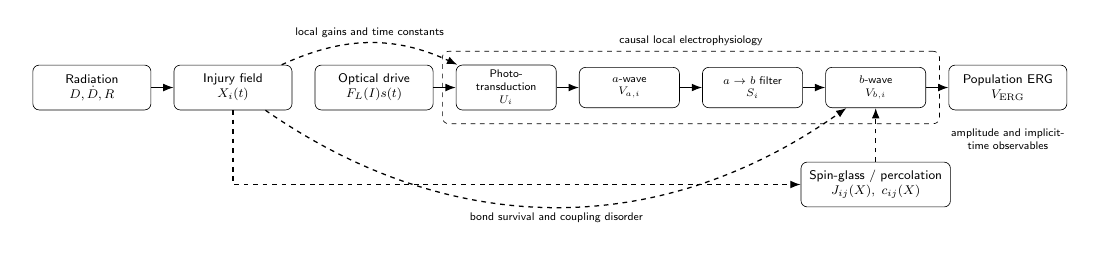}
\caption{Integrated causal model. Radiation creates a heterogeneous injury field $X_i$, which changes local phototransduction/electrophysiology and, in the network extension, the strength and survival of downstream inter-module couplings. Light intensity enters through $F_L(I)$ and propagates through $U_i\rightarrow V_{a,i}\rightarrow S_i\rightarrow V_{b,i}\rightarrow V_{\rm ERG}$. Dashed paths denote injury-dependent modulation and the optional spin-glass--percolation feedback.}
\label{fig:architecture}
\end{figure}

Figure~\ref{fig:architecture} summarizes the model. Light intensity first enters through a common saturating stimulus-response relation,
\begin{equation}
F_L(I)=\frac{I^{n_L}}{I^{n_L}+I_{50}^{n_L}},\qquad L_i(t;I)=L_0F_L(I)\phi_L(t),
\label{eq:hill}
\end{equation}
where $I$ is the optical light-flash intensity, $I_{50}$ is the half-saturation intensity, $n_L$ is a Hill/Naka--Rushton exponent, $L_0$ is an overall optical-drive normalization, and $\phi_L(t)$ is the normalized temporal profile of the optical light flash that evokes the ERG. Thus $\phi_L(t)$ describes the light stimulus and is distinct from the temporal structure of the X-ray FLASH irradiation pulse.

For retinal module $i$, the causal local pathway is
\begin{align}
\tau_{r,i}(X_i)\dot U_i+U_i &= g_{r,i}(X_i)L_i(t), \label{eq:u}\\
C_{a,i}(X_i)\dot V_{a,i}+\frac{V_{a,i}}{R_{a,i}(X_i)} &= -g_{a,i}(X_i)U_i, \label{eq:va}\\
\tau_{ab,i}(X_i)\dot S_i+S_i &= -V_{a,i}, \label{eq:s}\\
C_{b,i}(X_i)\dot V_{b,i}+\frac{V_{b,i}}{R_{b,i}(X_i)} &= g_{ab,i}(X_i)S_i+I_{\rm net,i}. \label{eq:vb}
\end{align}
All four states, $U_i, V_{a,i}, S_i$, and $V_{b,i}$, start from zero at $t=0$. The population readout is
\begin{equation}
\erg(t)=\alpha_a\frac{1}{N}\sum_{i=1}^N V_{a,i}(t)+\alpha_b\frac{1}{N}\sum_{i=1}^N V_{b,i}(t),\qquad N=200.
\label{eq:erg}
\end{equation}
This construction guarantees $\erg(0)=0$. It also yields a finite-rise a-wave and a dynamically delayed b-wave.

\subsection{Hodgkin--Huxley and retinal ion-channel basis of the rate equations}
The local equations in Eqs.~\eqref{eq:u}--\eqref{eq:vb} are not arbitrary phenomenological RC circuits. They are coarse-grained reductions of the same membrane-current balance that underlies the Hodgkin--Huxley formalism \cite{Hodgkin1952}. For a membrane patch with ionic pathways indexed by $k$,
\begin{equation}
C_m\frac{dV}{dt}
=-\sum_k g_k(V,t)\bigl(V-E_k\bigr)+I_{\rm drive}(t),
\label{eq:hh_generic}
\end{equation}
where $C_m$ is membrane capacitance, $g_k$ denotes the instantaneous conductance of an ion-channel pathway, and $E_k$ is its reversal potential. Linearization around a physiological operating point and coarse-graining of the channel-gating variables gives
\begin{equation}
C_{\rm eff}\frac{d\,\delta V}{dt}+G_{\rm eff}\,\delta V
\simeq I_{\rm drive}(t),
\qquad R_{\rm eff}=G_{\rm eff}^{-1},\qquad
\tau=R_{\rm eff}C_{\rm eff},
\label{eq:hh_rc_reduction}
\end{equation}
which has the same mathematical structure as the $a$- and $b$-compartment equations used here. Because photoreceptors and most bipolar cells signal primarily through graded changes in membrane potential, this Hodgkin--Huxley-to-RC reduction is a natural mesoscopic approximation; the model does not claim that retinal tissue consists of literal passive resistors and capacitors. 

The ion-channel and synaptic biology represented by the causal chain can be stated explicitly. In the photoreceptor stage, light activates rhodopsin, transducin, and PDE6, lowers cGMP, and closes CNG channels carrying the inward Na$^+$/Ca$^{2+}$ dark current; the resulting reduction of inward current, opposed by K$^+$ efflux, hyperpolarizes rods and cones \cite{FuYau2007,VanHook2019}. This molecular cascade is coarse-grained into $U_i$ and the photoreceptor-dominated state $V_{a,i}$. At the photoreceptor ribbon synapse, hyperpolarization reduces activation of L-type Ca$_V$1.4 channels and therefore reduces glutamate release \cite{Heidelberger2005,VanHook2019}. The variable $S_i$ represents this finite photoreceptor-to-bipolar signaling step. In ON bipolar cells, reduced glutamate decreases mGluR6/G$_o$ signaling, relieves inhibition of TRPM1, and opens a nonselective cation conductance carrying Na$^+$ and Ca$^{2+}$, producing the depolarizing response that contributes strongly to the ERG b-wave \cite{Masu1995,Morgans2009,Martemyanov2017}. The downstream state $V_{b,i}$ coarse-grains this TRPM1-mediated ON-bipolar response together with additional K$^+$ and Cl$^-$ conductances and extracellular/M\"uller-cell current contributions that shape the population b-wave \cite{Frishman2006,VanHook2019}.

Thus, the present model includes these channel pathways at an effective or lumped level rather than as separate microscopic state variables. CNG, photoreceptor K$^+$, Ca$_V$1.4, and TRPM1 currents are biologically represented in the causal stages, whereas individual Hodgkin--Huxley gating variables, explicit Na$_V$ action-potential generation, OFF-bipolar ionotropic glutamate-receptor pathways, amacrine-cell channel kinetics, and ganglion-cell spiking are not separately resolved in the current $N=200$ implementation. This distinction preserves a direct biophysical interpretation while keeping the number of fitted parameters compatible with the available ERG observables.

\subsection{Radiation injury and delayed follow-up bridge}
The underlying dynamic injury variable satisfies
\begin{equation}
\frac{dX_i}{dt}=k_{\rm inj,i}S_R(R)\dot D_i(t)-k_{\rm rec,i}X_i(t).
\label{eq:injury}
\end{equation}
For a square irradiation pulse of total dose $D$ delivered at dose rate $R$, duration $T=D/R$, the acute endpoint has the form
\begin{equation}
X^*(D,R)=k_{\rm inj}S_R D\frac{1-e^{-k_{\rm rec}D/R}}{k_{\rm rec}D/R}.
\label{eq:endpoint}
\end{equation}
Because the ERG measurements are delayed relative to irradiation, the complete fit applies one global follow-up scaling parameter to bridge the acute injury score to the follow-up state. This is deliberately parsimonious because a single follow-up time cannot identify a detailed chronic recovery law.

The notation in Eqs.~\eqref{eq:hill}--\eqref{eq:endpoint} is summarized in Table~\ref{tab:parameters}. A dot denotes a time derivative and $i$ labels a coarse-grained retinal functional module. The states $X_i$, $U_i$, $V_{a,i}$, $S_i$, and $V_{b,i}$ are calculated dynamically. The effective $R$ and $C$ quantities are Hodgkin--Huxley-inspired lumped conductance/capacitance parameters rather than literal single-cell resistors and capacitors, and $X_i$ is a latent injury coordinate rather than a unique molecular lesion.

\begingroup
\scriptsize
\renewcommand{\arraystretch}{0.94}
\setlength{\tabcolsep}{2pt}
\setlength{\LTleft}{0pt}
\setlength{\LTright}{0pt}
\begin{longtable}{@{}p{0.14\linewidth}p{0.39\linewidth}p{0.17\linewidth}p{0.25\linewidth}@{}}
\multicolumn{4}{@{}p{0.95\linewidth}@{}}{\label{tab:parameters}\small\textbf{Table \thetable:} Definitions, biological interpretation, and numerical status of the principal variables and parameters entering Eqs.~\eqref{eq:hill}--\eqref{eq:endpoint}, the associated network feedback, and the reduced-amplitude benchmark. The local electrophysiological quantities are Hodgkin--Huxley-inspired lumped variables that coarse-grain retinal ion-channel and synaptic pathways. Unless explicitly labeled ``reduced benchmark,'' ``fitted'' denotes values reported by the full M2 amplitude fit. ``Working baseline'' denotes fixed or normalized values used to define model time and response scales; these are not claimed as experimentally measured single-channel constants.}\\[4pt]
\toprule
Symbol & Definition / role & Units or type & Numerical value or status \\
\midrule
\endfirsthead
\multicolumn{4}{@{}l}{\tablename~\thetable\ (continued)}\\
\toprule
Symbol & Definition / role & Units or type & Numerical value or status \\
\midrule
\endhead
\multicolumn{4}{@{}l}{\textit{Optical input and dynamical states}}\\[-1pt]
$i$ & retinal functional-module index & index & $i=1,\ldots,N$ \\
$N$ & number of coarse-grained retinal modules & dimensionless & $200$ \\
$I$ & optical light-flash intensity used to evoke the ERG & cd\,s\,m$^{-2}$ & experimental input \\
$F_L(I)$ & normalized saturating light-response function & dimensionless & calculated from Eq.~\eqref{eq:hill} \\
$I_{50}$ & half-saturation optical intensity & cd\,s\,m$^{-2}$ & $3.95\times10^{-2}$, fitted (full M2) \\
$n_L$ & Hill/Naka--Rushton exponent & dimensionless & $0.327$, fitted (full M2) \\
$L_0$ & overall optical-drive normalization & normalized & $1$, working normalization \\
$\phi_L(t)$ & temporal profile of the optical light flash; not the X-ray FLASH pulse & normalized function & 3-ms square pulse in the working causal implementation \\
$X_i(t)$ & effective local radiation-injury state & latent/model units & calculated from Eq.~\eqref{eq:injury}; $X_i=0$ for control \\
$U_i(t)$ & phototransduction state: rhodopsin/transducin/PDE6/cGMP control of CNG Na$^+$/Ca$^{2+}$ current & normalized state & dynamically calculated \\
$V_{a,i}(t)$ & photoreceptor membrane/a-wave state: net CNG inward current and opposing K$^+$ conductance & model voltage & dynamically calculated \\
$S_i(t)$ & photoreceptor-to-ON-bipolar signaling state: Ca$_V$1.4-regulated glutamate release and mGluR6/G$_o$ cascade & model state & dynamically calculated \\
$V_{b,i}(t)$ & ON-bipolar/post-photoreceptor state: TRPM1 cation current plus lumped K$^+$/Cl$^-$ and extracellular contributions & model voltage & dynamically calculated \\
$I_{\rm net,i}$ & collective input from neighboring b-layer modules & model drive & $0$ in M0; calculated from Eq.~\eqref{eq:feedback} in M1/M2 \\
$\alpha_a,\alpha_b$ & population weights converting mean a- and b-layer states to total ERG & scale factors & output-normalization parameters; not independently interpreted as biological constants \\
\midrule
\multicolumn{4}{@{}l}{\textit{Local causal kinetic scales}}\\[-1pt]
$\tau_{r,i}$ & phototransduction rise/processing time & time & injury dependent; $\tau_{r0}=2.5$ ms working baseline \\
$g_{r,i}$ & phototransduction gain & normalized gain & $g_{r0}=1.0$ working baseline \\
$R_{a,i},C_{a,i}$ & HH/RC reduction of effective photoreceptor conductance and capacitance; $\tau_{a,i}=R_{a,i}C_{a,i}$ & normalized $R$; time-like $C$ & $R_{a0}=1.0$, $C_{a0}=0.012$ s working baselines \\
$g_{a,i}$ & drive from $U_i$ to the negative a-wave compartment & model gain & $g_{a0}=4.0$ working baseline \\
$\tau_{ab,i}$ & photoreceptor-to-post-photoreceptor signaling time & time & $\tau_{ab0}=8.0$ ms working baseline \\
$g_{ab,i}$ & coupling from $S_i$ to the b-wave compartment & model gain & $g_{ab0}=2.30$ working baseline \\
$R_{b,i},C_{b,i}$ & HH/RC reduction of effective ON-bipolar/downstream conductance and capacitance; $\tau_{b,i}=R_{b,i}C_{b,i}$ & normalized $R$; time-like $C$ & $R_{b0}=1.0$, $C_{b0}=0.045$ s working baselines \\
\midrule
\multicolumn{4}{@{}l}{\textit{Radiation-injury variables and fitted bridge}}\\[-1pt]
$D$ & total ocular radiation dose & Gy & 21 or 26 Gy for irradiated groups \\
$R$ & irradiation dose rate for the square-pulse representation & Gy\,s$^{-1}$ & 52.6 (FLASH) or 1.1 (CONV) \\
$\dot D_i(t)$ & local instantaneous dose rate & Gy\,s$^{-1}$ & prescribed radiation input \\
$S_R(R)$ & dose-rate-condition injury modifier & dimensionless & $S_{\rm CONV}=1$ reference; $S_{\rm FLASH}=0.507$ fitted \\
$k_{\rm inj,i}$ & effective injury-induction coefficient & model scale & $k_{\rm inj}=0.20$ working baseline \\
$k_{\rm rec,i}$ & effective recovery coefficient & s$^{-1}$ & $k_{\rm rec}=0.020$ s$^{-1}$ working baseline \\
$T=D/R$ & radiation delivery duration in the square-pulse idealization & s & determined by $D$ and $R$ \\
$f_{\rm FU}$ & global acute-to-follow-up injury scale used for delayed ERG fitting & dimensionless & $0.487$, fitted \\
\midrule
\multicolumn{4}{@{}l}{\textit{Network and spin-diagnostic quantities}}\\[-1pt]
$s_i$ & coarse-grained Ising state of one retinal module (a local neuronal/cellular population), not a single neuron or magnetic spin & binary diagnostic variable & $+1$: active/responsive; $-1$: inactive/suppressed; diagnostic only \\
$\eta$ & continuous network-feedback strength & model scale & $0.0141$, fitted (M2) \\
$V_s$ & soft-spin voltage scale in $\tanh(V_{b,j}/V_s)$ & model voltage & $0.10$, working baseline \\
$z_0$ & target mean degree of the baseline random graph & dimensionless & $7$ \\
$\alpha_J$ & injury attenuation of coherent coupling & dimensionless & $0$, fitted (M2; weakly identifiable) \\
$\alpha_p$ & injury sensitivity of bond survival & dimensionless & $0.718$, fitted (M2) \\
$\kappa_\sigma$ & injury-induced increase of coupling disorder & dimensionless & $1.50$, fitted (M2; weakly identifiable) \\
\midrule
\multicolumn{4}{@{}l}{\textit{Reduced-amplitude benchmark parameters (distinct from M0--M2)}}\\[-1pt]
$I_{50}^{\rm red},\,n_L^{\rm red}$ & half-saturation intensity and Hill/Naka--Rushton exponent in the reduced benchmark & cd\,s\,m$^{-2}$; dimensionless & $0.3459$; $0.9383$, fitted \\
$\gamma_a^{\rm red},\,\gamma_b^{\rm red}$ & effective injury-sensitivity coefficients for reduced a- and b-wave gains & dimensionless & $0.3751$; $0.3300$, fitted \\
$\rho,\,q_D$ & CONV/FLASH injury ratio at 26 Gy and phenomenological dose exponent in $X_{\rm red}$ & dimensionless & $3.3848$; $6.4209$, fitted \\
$B_{w,c},\,A_{w,c}$ & control floor and dynamic range for wave $w\in\{a,b\}$ and control cohort $c\in\{F,C\}$ & $\mu$V & $B_{a,F}=19.304$, $A_{a,F}=51.748$; $B_{a,C}=23.169$, $A_{a,C}=77.279$; $B_{b,F}=96.226$, $A_{b,F}=122.085$; $B_{b,C}=126.330$, $A_{b,C}=147.952$ \\
$G_w(D,R)$ & derived gain-survival factor, $\exp[-\gamma_w^{\rm red}X_{\rm red}(D,R)]$ & dimensionless & derived from fitted parameters; not independently fitted \\
\bottomrule
\end{longtable}
\endgroup

Table~\ref{tab:parameters} separates dynamical states, working baseline scales, and fitted quantities. The full-model fit reports $I_{50}$, $n_L$, $S_{\rm FLASH}$, the follow-up scale, and the listed network parameters, whereas the reduced-amplitude benchmark has its own explicitly labeled parameter set $I_{50}^{\rm red}$, $n_L^{\rm red}$, $\gamma_{a,b}^{\rm red}$, $\rho$, $q_D$, $B_{w,c}$, and $A_{w,c}$. These benchmark parameters should not be identified with the corresponding parameters of M0--M2. The local baselines set normalization and characteristic time scales and should be interpreted as model settings rather than experimentally established membrane or single-channel constants. In particular, the table identifies which ion-channel pathways are represented by each coarse-grained state, while Eqs.~\eqref{eq:hh_generic} and \eqref{eq:hh_rc_reduction} explain how their combined conductances motivate the RC form. The present amplitude data constrain effective population behavior more strongly than individual microscopic kinetic parameters.

\subsection{Injury dependence of local electrophysiology}
The theory permits injury to alter amplitude and timing independently. Representative parameterizations are
\begin{align}
\tau_{r,i}(X_i)&=\tau_{r0,i}(1+\alpha_{\tau r}X_i), & g_{r,i}(X_i)&=g_{r0,i}e^{-\gamma_rX_i},\\
R_{a,i}(X_i)&=R_{a0,i}(1+\alpha_{R_a}X_i), & C_{a,i}(X_i)&=C_{a0,i}(1+\alpha_{C_a}X_i),\\
R_{b,i}(X_i)&=R_{b0,i}(1+\alpha_{R_b}X_i), & C_{b,i}(X_i)&=C_{b0,i}(1+\alpha_{C_b}X_i),\\
g_{a,i}(X_i)&=g_{a0,i}e^{-\gamma_aX_i}, & g_{ab,i}(X_i)&=g_{ab0,i}e^{-\gamma_{ab}X_i},\\
\tau_{ab,i}(X_i)&=\tau_{ab0,i}(1+\alpha_{\tau ab}X_i).&&
\end{align}
The fitted data therefore constrain population-level coefficients while fixed baseline heterogeneity generates module-to-module variability.

\subsection{Spin-glass and percolation extension}
For pair injury $X_{ij}=(X_i+X_j)/2$, surviving effective couplings are modeled as
\begin{equation}
J_{ij}(X_{ij})=\frac{\bar J_0}{z_0}e^{-\alpha_JX_{ij}}
+\frac{\sigma_0+\kappa_\sigma X_{ij}}{\sqrt{z_0}}\xi_{ij},\qquad \xi_{ij}\sim\mathcal N(0,1),
\label{eq:J}
\end{equation}
whereas functional bond survival is independent of coupling sign,
\begin{equation}
\Pr(c_{ij}=1\mid X_{ij})=e^{-\alpha_pX_{ij}}.
\label{eq:bond}
\end{equation}
Here, $c_{ij}\in\{0,1\}$ is the bond-survival variable, with $c_{ij}=1$ for a preserved functional connection and $c_{ij}=0$ for a lost connection. The form
$\Pr(c_{ij}=1\mid X_{ij})=\exp(-\alpha_p X_{ij})$
therefore represents an injury-dependent bond-survival probability: when $X_{ij}=0$ the connection survives with probability one, whereas increasing injury progressively reduces the probability of retaining that bond, at a rate controlled by $\alpha_p$.
In the limit $\alpha_p=0$, one obtains
$\Pr(c_{ij}=1\mid X_{ij})=1$ for all $X_{ij}$, so no radiation-induced bond deletion occurs and the network topology remains fixed.
In the opposite limit $\alpha_p\rightarrow\infty$, the bond-survival
probability approaches zero for any finite positive injury,
$X_{ij}>0$, while remaining equal to unity at $X_{ij}=0$, i.e., 
$\Pr(c_{ij}=1\mid X_{ij}=0)=1$ for any $\alpha_p$ including $\alpha_p\rightarrow\infty$. Thus, $\alpha_p$ controls the sharpness of radiation-induced connectivity loss: $\alpha_p=0$ corresponds to no bond deletion, whereas very large $\alpha_p$ produces an effectively abrupt loss of any bond exposed to nonzero injury.
The continuous b-layer feedback is
\begin{equation}
I_{\rm net,i}=\eta\sum_jc_{ij}J_{ij}\tanh\!\left(\frac{V_{b,j}}{V_s}\right).
\label{eq:feedback}
\end{equation}

For the statistical-mechanical diagnostics, each coarse-grained retinal module $i$ is additionally represented by an Ising spin variable
\begin{equation}
s_i\in\{-1,+1\}.
\label{eq:spinstate}
\end{equation}
The spin is \emph{not} a magnetic moment and does not represent the firing state of one literal neuron. Rather, $s_i$ represents the aggregate functional state of a mesoscopic retinal module, i.e., a local collection of photoreceptors, bipolar cells, interneurons, and supporting retinal cells whose activity has been coarse-grained into one binary diagnostic variable. We define $s_i=+1$ as a functionally active or responsive module: the local population participates coherently in stimulus-evoked retinal signaling and is aligned with the locally favored network state. We define $s_i=-1$ as a functionally inactive, suppressed, or oppositely aligned module: its net participation in the coordinated response is reduced relative to that reference state. Because much of retinal signaling is graded rather than action-potential based, ``active'' and ``inactive'' should not be interpreted simply as spiking versus nonspiking. Moreover, $s_i=-1$ does not imply that the cells are dead or structurally absent; it denotes a functional state within the binary coarse-grained projection.

The hyperbolic-tangent transfer function is chosen to provide a bounded
continuous analogue of the binary Ising spin. In mean-field Ising theory,
the local average spin has the familiar form
$m_i=\tanh(\beta h_i^{\rm eff})$, where $h_i^{\rm eff}$ is the effective
local field. Analogously, in the present continuous retinal model,
$\tanh(V_{b,j}/V_s)$ maps the continuous downstream voltage $V_{b,j}$ onto
a dimensionless signed activity variable in the interval $[-1,1]$.
For $|V_{b,j}|\gg V_s$, the function approaches the asymptotic states
$\pm1$, corresponding to the two coarse-grained functional states used
in the associated Ising description, whereas for $|V_{b,j}|\ll V_s$,
$\tanh(V_{b,j}/V_s)\simeq V_{b,j}/V_s$ and the network interaction is
approximately linear. Thus, the $\tanh$ function provides a
``soft-spin'' bridge between the continuous ERG voltage dynamics and
the binary spin representation, while also preventing unbounded
network feedback. The scale $V_s$ controls how rapidly the continuous
activity approaches the two saturated spin-like states.

With this interpretation, the spin part of the model is summarized by the diagnostic Ising Hamiltonian
\begin{equation}
H_{\rm spin}=-\sum_{i<j}c_{ij}J_{ij}s_is_j-\sum_i h_i s_i,
\label{eq:spinham}
\end{equation}
where $c_{ij}\in\{0,1\}$ indicates whether a functional bond survives, $J_{ij}$ is the signed effective interaction between the two modules, and $h_i$ is a local bias toward one functional state. A positive $J_{ij}$ favors modules $i$ and $j$ occupying the same binary state, whereas a negative $J_{ij}$ favors opposite states. Competition among positive and negative couplings on the surviving network generates frustration and glass-like disorder. Heat-bath/Glauber spin updates are used only to sample these collective diagnostic states; they do not replace the continuous $U_i\rightarrow V_{a,i}\rightarrow S_i\rightarrow V_{b,i}$ dynamics that generate the modeled ERG waveform.

The resulting order parameters are
\begin{equation}
m=\frac{1}{N}\sum_i\langle s_i\rangle,\qquad
q_{\rm EA}=\frac{1}{N}\sum_i\langle s_i\rangle^2,\qquad
q_{\rm SG}=q_{\rm EA}-m^2,
\label{eq:spinorders}
\end{equation}
where $m$ measures coherent module-level order, $q_{\rm EA}$ measures persistent local ordering even when the global mean is small, and $q_{\rm SG}$ isolates the excess glassy component. The frustration fraction counts unsatisfied surviving interactions, for which $J_{ij}s_is_j<0$, while the surviving-bond fraction $p$ quantifies structural/functional connectivity. Thus the spin observables describe collective organization of coarse-grained retinal populations rather than the activity of individual neurons.

The Hamiltonian $H_{\rm spin}$ should be interpreted as an effective statistical-physics cost function rather than as the literal biochemical or electrical energy of the retina. Its minimum defines the collective spin configuration that is most compatible with the surviving network bonds, the effective couplings $J_{ij}$, and the local bias fields $h_i$. For predominantly positive couplings and positive bias, the low-energy state is largely aligned, $s_i\simeq +1$, corresponding in the present coarse-grained interpretation to a retinal network in which most functional modules---each representing a local collection of retinal neurons and supporting cells rather than a single neuron---remain coherently responsive to the optical stimulus, resembling the state of the healthy retina. As radiation injury weakens couplings, introduces sign disorder, and removes bonds, the interactions can become mutually incompatible, producing frustration and multiple competing low-energy or metastable configurations. These states represent heterogeneous persistent patterns of collective retinal activity in which different modules can remain preferentially active or suppressed even though global coherence is reduced. Thus, the ground state provides a useful reference for the maximally compatible collective retinal configuration, while the finite-noise Glauber dynamics explores excited and metastable states of the same effective energy landscape. In this sense, a transition from a strongly aligned low-energy state toward many competing minima provides a statistical-mechanical description of radiation-induced loss of collective retinal organization.

The independent finite-size percolation scan locates $\pc$ from the peak of finite-cluster susceptibility; for the chosen $N=200$, mean-degree-$\simeq7$ ensemble, $\pc\simeq0.17$. This is a topology-dependent model scale, not a universal biological threshold.

\subsection{Two fitting approaches}
The first fitting approach is a reduced amplitude benchmark, a compact reference model for the measured a- and b-wave amplitudes. It uses the same Hill/Naka--Rushton functional form as Eq.~\eqref{eq:hill}, but with benchmark-specific parameters $I_{50}^{\rm red}$ and $n_L^{\rm red}$, together with the phenomenological radiation-injury law
\begin{equation}
X_{\rm red}(D,R)
=
\rho_R
\left(
\frac{D}{26\,\mathrm{Gy}}
\right)^{q_D},
\qquad
\rho_{\rm FLASH}=1,
\qquad
\rho_{\rm CONV}=\rho .
\label{eq:reduced_injury}
\end{equation}
For each ERG component $w\in\{a,b\}$, we define the dimensionless radiation-dependent gain-survival factor
\begin{equation}
G_w(D,R)
=
\exp\!\left[-\gamma_w^{\rm red} X_{\rm red}(D,R)\right],
\label{eq:gain_survival}
\end{equation}
so that
\begin{equation}
G_a(D,R)=\exp[-\gamma_a^{\rm red} X_{\rm red}(D,R)],
\qquad
G_b(D,R)=\exp[-\gamma_b^{\rm red} X_{\rm red}(D,R)].
\label{eq:gain_survival_ab}
\end{equation}
Here, $\gamma_a^{\rm red}$ and $\gamma_b^{\rm red}$ characterize the effective injury sensitivity of the reduced a- and b-wave amplitudes, respectively. The reduced amplitude model can therefore be written as
\begin{equation}
A^{\rm red}_{w,c}(I;D,R)
=
\left[
B_{w,c}
+
A_{w,c}F_L^{\rm red}(I)
\right]
G_w(D,R),
\label{eq:reduced_amplitude}
\end{equation}
with
\begin{equation}
F_L^{\rm red}(I)=
\frac{I^{n_L^{\rm red}}}
{I^{n_L^{\rm red}}+(I_{50}^{\rm red})^{n_L^{\rm red}}}.
\label{eq:reduced_hill}
\end{equation}
Here $B_{w,c}$ and $A_{w,c}$ are the control floor and dynamic-range parameters for wave $w$ and control cohort $c$. Thus $G_w=1$ corresponds to the unattenuated control response, while $G_w<1$ represents radiation-induced suppression. The superscript ``red'' emphasizes that $X_{\rm red}$, $I_{50}^{\rm red}$, $n_L^{\rm red}$, and $\gamma_w^{\rm red}$ belong to the phenomenological reduced benchmark and are distinct from the dynamic injury state $X_i(t)$ and the local kinetic parameters of the M0--M2 causal framework.

The experimental amplitude-versus-intensity responses were fitted simultaneously with an unweighted residual objective. This reduced formulation provides a compact reference for the stimulus-response and radiation-injury dependence without attempting to resolve the full network dynamics.

The second fitting approach uses the complete causal model and a common fixed realization of baseline heterogeneity and quenched graph disorder during optimization. Three nested realizations were compared: M0 sets $\eta=0$, in Eq. (\ref{eq:feedback}), hence $I_{\rm net,i}=0$ and therefore contains causal local electrophysiology without network feedback.
Thus, it is the local-only version of the full multiscale causal model and contains the optical input, injury dynamics, and local electrophysiologic/RC response, but sets the collective network feedback to zero.
M1 adds collective network coupling; thus, it includes disordered $J_{ij}$ but sets $\alpha_p=0$ in Eq. (\ref{eq:bond}) where no radiation-induced bond deletion occurs
and the network topology remains fixed.
M2 adds the full spin-glass/percolation-dependent network terms; thus, it includes both disordered coupling and injury-dependent bond survival, as shown in the following equation in the box 
\begin{equation}
\boxed{
\text{Reduced-amplitude benchmark}
\quad \text{vs.} \quad
\underbrace{
\mathrm{M0}\rightarrow\mathrm{M1}\rightarrow\mathrm{M2}
}_{\text{full causal multiscale framework}}
}
\end{equation}
To balance the different characteristic scales of the a- and b-wave amplitudes, a common scale-balanced objective was used,
\begin{equation}
\mathcal L^*=\sum_{g,k}\left[\frac{A^{(g)}_{a,\rm mod}-A^{(g)}_{a,\rm exp}}{30\,\mu\mathrm V}\right]^2+
\sum_{g,k}\left[\frac{A^{(g)}_{b,\rm mod}-A^{(g)}_{b,\rm exp}}{65\,\mu\mathrm V}\right]^2.
\label{eq:objective}
\end{equation}
The resulting $\mathrm{AIC}^*$ and $\mathrm{BIC}^*$ are used only for relative comparison among the nested realizations. The asterisk emphasizes that these quantities are screening metrics defined from the common scale-balanced objective rather than a full eye-level likelihood.

To translate the reduced benchmark into a prospective dose-response prediction, we define an exploratory functional-preservation threshold $D^*_{w,r}$ as the dose at which the retained gain for ERG component $w\in\{a,b\}$ and irradiation mode $r\in\{\mathrm{FLASH},\mathrm{CONV}\}$ falls to a specified fraction $f_{\rm crit}$ of the control gain,
\begin{equation}
G_w(D^*_{w,r},r)=f_{\rm crit}.
\label{eq:functional_threshold_def}
\end{equation}
Using Eq.~\eqref{eq:gain_survival} and $\rho_{\rm FLASH}=1$, $\rho_{\rm CONV}=\rho$, this gives
\begin{equation}
D^*_{w,r}(f_{\rm crit})
=26\,\mathrm{Gy}\left[
\frac{-\ln f_{\rm crit}}{\gamma_w^{\rm red}\rho_r}
\right]^{1/q_D}.
\label{eq:functional_threshold}
\end{equation}
We define the corresponding model-based functional dose-modifying factor as
\begin{equation}
\mathrm{DMF}_{w}(f_{\rm crit})
=\frac{D^*_{w,\rm FLASH}}{D^*_{w,\rm CONV}}
=\rho^{1/q_D}.
\label{eq:functional_dmf}
\end{equation}
For prospective illustration we use $f_{\rm crit}=0.50$, i.e., 50\% retention of the control dynamic gain. This criterion is a transparent modeling convention rather than an experimentally established retinal-tolerance or clinical prescription threshold. Because the calibration data contain only 21- and 26-Gy irradiated groups, predictions above 26 Gy constitute extrapolation and require direct experimental validation.

\section{Results}
\subsection{Reduced amplitude benchmark}
The reduced model achieved a high level of agreement with the experimental group-mean amplitudes, producing $R^2=0.9721$, RMSE $=11.94~\mu$V, and MAE $=9.36~\mu$V. The fitted benchmark stimulus parameters were $I_{50}^{\rm red}=0.3459$ cd s m$^{-2}$ and $n_L^{\rm red}=0.9383$. The fitted reduced injury-gain coefficients were $\gamma_a^{\rm red}=0.3751$ and $\gamma_b^{\rm red}=0.3300$, with an effective CONV/FLASH injury ratio $\rho=3.3848$ and a steep phenomenological dose exponent $q_D=6.4209$. The corresponding $X_{\rm red}$ injury scores were 0.254, 1.000, 0.859, and 3.385 for FLASH21, FLASH26, CONV21, and CONV26. FLASH21 and FLASH26 denote FLASH irradiation at 21 and 26 Gy, whereas CONV21 and CONV26 denote conventional-dose-rate irradiation at the same doses. These quantities summarize the amplitude data effectively but should be interpreted as phenomenological parameters rather than uniquely identified biological constants.

\begin{figure}[t]
\centering
\includegraphics[width=0.96\linewidth]{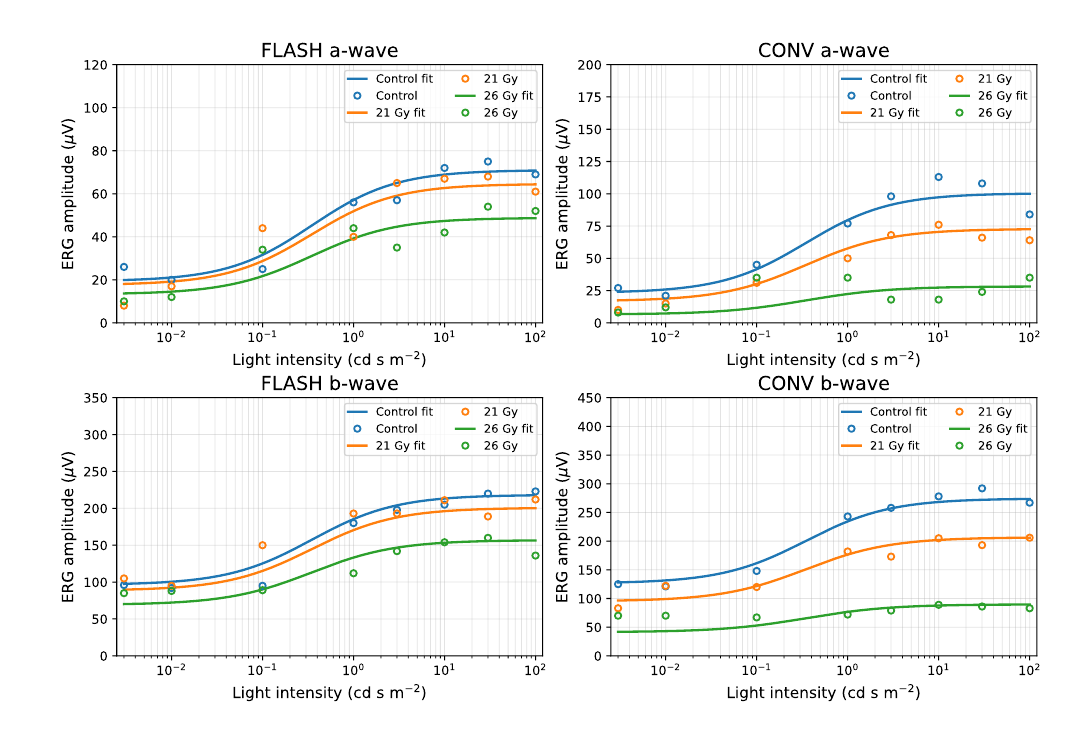}
\caption{Reduced amplitude benchmark for the FLASH and CONV a- and b-wave intensity-response data. Open symbols denote the experimental group-mean ERG amplitudes and solid curves denote the fitted reduced-model responses for Control, 21 Gy, and 26 Gy. This phenomenological benchmark provides a compact reference for stimulus-response and radiation-injury dependence and is distinct from the nested M0/M1/M2 mechanistic comparison.}
\label{fig:reducedfit}
\end{figure}

Using the reduced-benchmark parameters summarized in Table~\ref{tab:parameters}, the gain-survival factors obtained from Eq.~\eqref{eq:gain_survival} were $G_a=(0.909,0.687,0.725,0.281)$ and $G_b=(0.920,0.719,0.753,0.327)$ for FLASH21, FLASH26, CONV21, and CONV26, respectively. Thus the benchmark captures strong preservation after FLASH and severe suppression after 26-Gy CONV while maintaining a common light-response nonlinearity. Figure~\ref{fig:reducedfit} shows the corresponding reduced-model fits to the experimental group-mean amplitudes.

\subsection{Full causal multiscale fit}
\begin{figure}[t]
\centering
\includegraphics[width=0.94\linewidth]{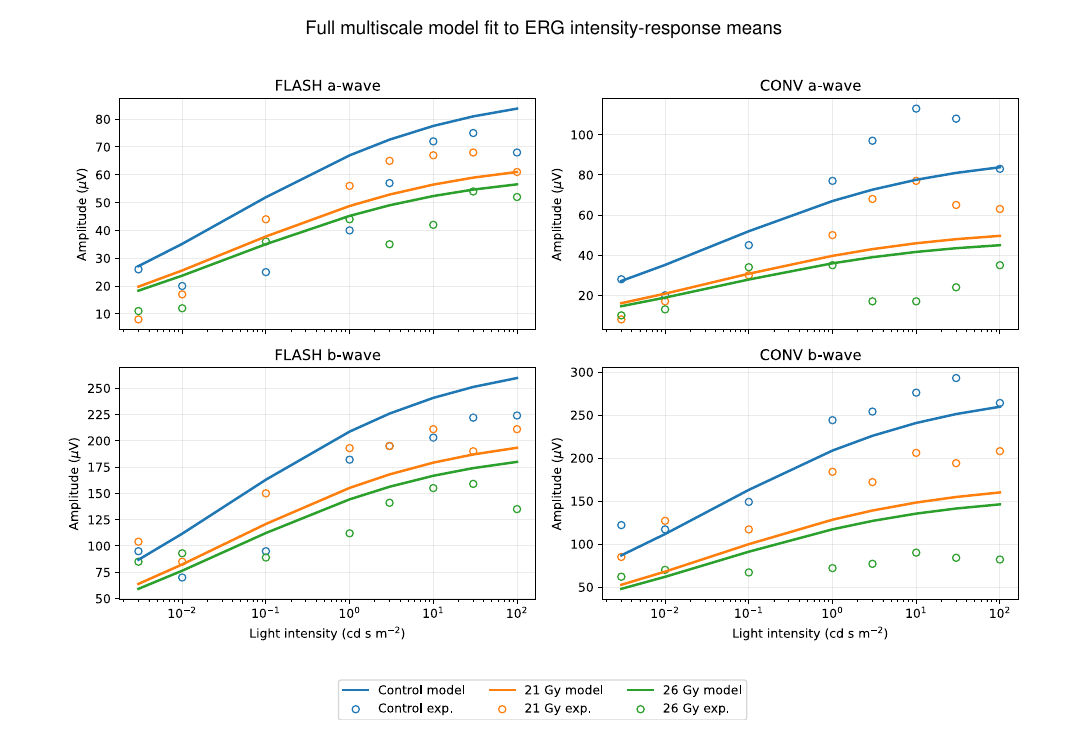}
\caption{Experimental ERG intensity-response means (open symbols) and predictions from the fitted full M2 $N=200$ model (solid curves). The same model control curve is compared with the control measurements shown in the FLASH and CONV panels.}
\label{fig:fullfit}
\end{figure}

In contrast to the reduced benchmark in Fig.~\ref{fig:reducedfit}, Figure~\ref{fig:fullfit} shows the full M2 prediction against the experimental intensity-response means. The principal fitted values are summarized together with the definitions and working model scales in Table~\ref{tab:parameters}. The full model fitted $I_{50}=3.95\times10^{-2}$ cd s m$^{-2}$, $n_L=0.327$, $S_{\rm FLASH}=0.507$, and a global follow-up injury scale of 0.487. Effective mean injury scores were approximately 1.03, 1.28, 1.70, and 2.02 for FLASH21, FLASH26, CONV21, and CONV26. The network parameters were $\eta=0.0141$, $\alpha_J=0$, $\alpha_p=0.718$, and $\kappa_\sigma=1.50$. The boundary tendencies of $\alpha_J$ and $\kappa_\sigma$ indicate limited identifiability; these fitted values should therefore not be interpreted mechanistically.

\subsection{Nested comparison and identifiability of network contributions}
\begin{figure}[t]
\centering
\includegraphics[width=0.72\linewidth]{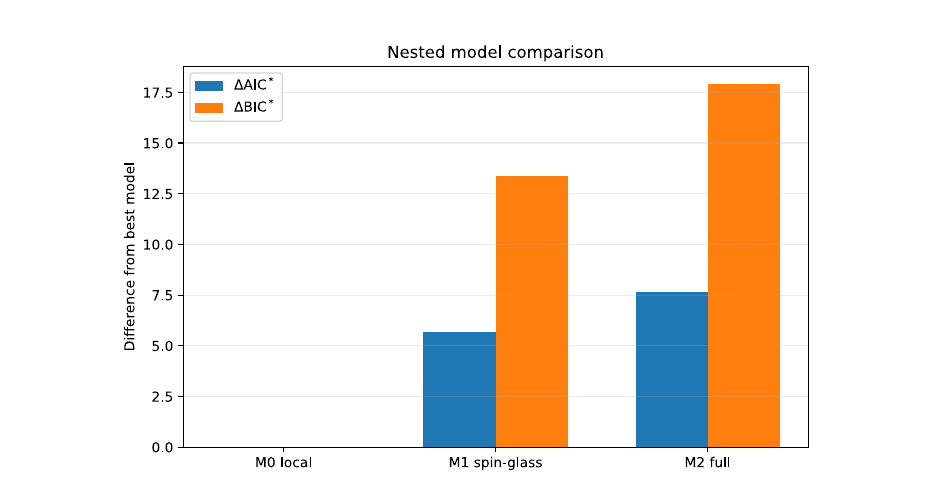}
\caption{Information-criterion screening of the nested M0/M1/M2 comparison. Lower $\Delta\mathrm{AIC}^*$ and $\Delta\mathrm{BIC}^*$ are preferred within this scale-balanced analysis. M0 has the lowest penalized screening values, while M1 and M2 provide closely related fits with additional network degrees of freedom, indicating that amplitude measurements alone do not independently resolve the network terms.}
\label{fig:modelselection}
\end{figure}

Figure~\ref{fig:modelselection} summarizes the nested mechanistic comparison. The M0, M1, and M2 realizations used $k=8$, 11, and 12 fitted parameters and yielded $\mathcal L^*=25.7721$, 25.6891, and 25.6756; RMSE $=27.2385$, 27.1188, and $27.1059~\mu$V; and $R^2=0.8546$, 0.8559, and 0.8560, respectively. Relative to M0, the screening differences were $\Delta\mathrm{AIC}^*/\Delta\mathrm{BIC}^*=5.69/13.38$ for M1 and $7.64/17.90$ for M2. Thus M1 and M2 lower $\mathcal L^*$ slightly relative to M0, with M2 improving RMSE by about $0.13~\mu$V and lowering the scale-balanced objective by about $0.37\%$, but the corresponding information-criterion screening values favor M0 for the present amplitude data. This result is interpreted as limited independent identifiability of the network terms from amplitude-versus-intensity measurements, rather than evidence that collective network effects are absent.

The apparently stronger global $R^2$ of the reduced benchmark should not be compared as though the two calculations were the same statistical model. The reduced fit is a flexible amplitude-level projection with separate control baselines for the two experimental panels and an unweighted objective, whereas the multiscale fit imposes the causal $N=200$ architecture, a common control response, and a hierarchical parameter set linking local dynamics to collective network behavior. The reduced fit therefore serves as a calibration benchmark, while the full M0/M1/M2 calculation tests how strongly the present observables distinguish local injury from collective disordered-network effects. Metrics are interpreted within each fitting formulation rather than compared through a common information criterion across the two formulations.

\subsection{Network predictions for experimental validation}
\begin{figure}[t]
\centering
\includegraphics[width=0.72\linewidth]{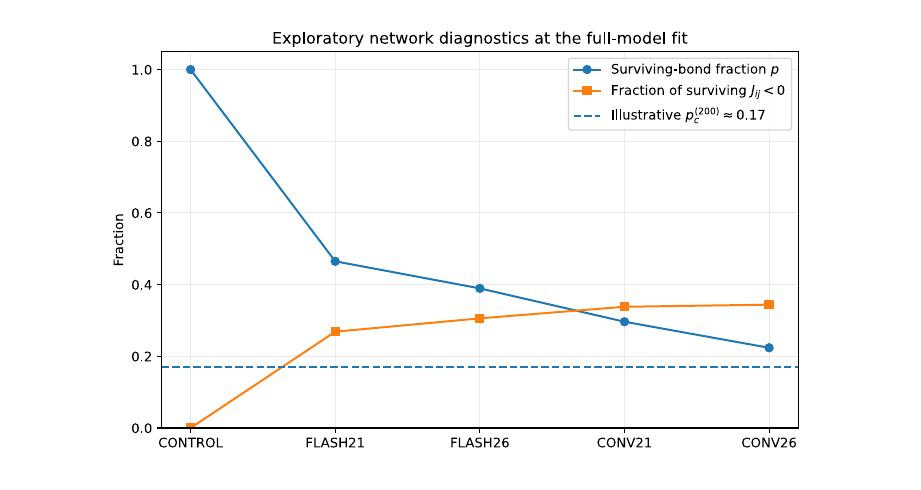}
\caption{Exploratory structural diagnostics from the representative full-model fit. Bond survival decreases and the fraction of negative surviving couplings increases with fitted injury. The dashed line is the independently calculated finite-size scale $\pc\approx0.17$. The curves are model predictions, not measured network quantities.}
\label{fig:network}
\end{figure}

The fitted M2 realization makes concrete predictions that extend beyond the information contained in ERG amplitude alone. The retained-bond fractions are approximately 1.00, 0.47, 0.39, 0.30, and 0.22 for Control, FLASH21, FLASH26, CONV21, and CONV26, respectively (Fig.~\ref{fig:network}). The fraction of negative surviving couplings rises with injury. Because these quantities are model-level network observables rather than direct measurements, they provide hypotheses for future validation using spatially resolved, temporal, structural, and correlation-sensitive experiments.

\subsection{Prospective prediction of the FLASH functional-preservation window}
The reduced benchmark can be used prospectively to identify doses at which the model predicts loss of a specified fraction of the control ERG gain. Using the exploratory criterion $f_{\rm crit}=0.50$ in Eq.~\eqref{eq:functional_threshold}, the predicted a-wave threshold is $D^*_{a,\rm FLASH}=28.6$ Gy for FLASH and $D^*_{a,\rm CONV}=23.7$ Gy for CONV. The corresponding b-wave thresholds are $D^*_{b,\rm FLASH}=29.2$ Gy and $D^*_{b,\rm CONV}=24.1$ Gy. Because the same reduced injury ratio $\rho$ and dose exponent $q_D$ govern both wave components, Eq.~\eqref{eq:functional_dmf} gives the same functional DMF for either component,
\begin{equation}
\mathrm{DMF}_{\rm eye}=\rho^{1/q_D}\simeq1.21.
\label{eq:dmf_numeric}
\end{equation}
A conservative combined ERG threshold, defined as the lower of the a- and b-wave thresholds, is therefore approximately 28.6 Gy for FLASH versus 23.7 Gy for CONV under the 50\% criterion. These values should be interpreted as prospective model predictions rather than measured tolerance doses. In particular, the predicted FLASH threshold lies above the highest irradiated dose used for calibration (26 Gy), so the model specifically predicts that additional dose-escalation measurements are needed to determine where FLASH functional preservation begins to fail. The resulting dose window provides a direct experimental test of the reduced injury law and, when combined with full-waveform and network-sensitive measurements, a route for testing whether the M0--M2 mechanisms change near the functional transition.

\section{Discussion}
This work develops a causal, multiscale framework for linking radiation exposure to retinal dysfunction measured by ERG. The model does not treat ERG amplitude as a purely descriptive endpoint; instead, it connects the physical radiation stimulus to local injury dynamics, downstream electrophysiologic response, and, in the full formulation, collective network behavior. In this context, the model provides both a compact phenomenological benchmark and a mechanistic extension that can generate testable hypotheses about how radiation alters retinal function. This combination is relevant to ocular radiation therapy because preservation of tumor control must ultimately be balanced against preservation of retinal function, and a quantitative model that links dose and dose rate to functional response can help define prospective experiments aimed at that balance.

An important result is that the reduced benchmark reproduces the observed intensity-response behavior and yields a prospective functional FLASH DMF of approximately 1.21 under the exploratory 50\% retained-gain criterion. This corresponds to a conservative combined threshold, defined as the lower of the a- and b-wave thresholds, of about 28.6 Gy for FLASH versus 23.7 Gy for CONV under the modeled single-fraction kV X-ray conditions. Although these threshold estimates are extrapolative and should not be interpreted as clinical tolerance doses, they demonstrate that the fitted response law can translate ERG measurements into a modality-sensitive dose prediction. This moves the analysis beyond retrospective curve fitting toward a predictive framework for comparing radiation-delivery modes in the eye. The underlying experiment used healthy mouse eyes, did not test tumor control, and included irradiated doses only through 26 Gy; therefore, the predicted FLASH threshold remains a prospective hypothesis requiring direct dose-escalation validation.

The nested mechanistic comparison provides a second important finding. The full M0$\rightarrow$M1$\rightarrow$M2 framework improves the residual objective only modestly relative to the local-only model, indicating that amplitude-versus-intensity data alone do not independently identify the network parameters. However, this should not be interpreted as evidence against collective retinal-network injury. Rather, it shows that ERG amplitude is a low-dimensional observable that can be generated by multiple underlying mechanisms. The collective terms in M1 and M2, represented by the spin-glass/percolation layer, remain useful because they predict network-level observables that are not accessible to the local model, including coupling-sign disorder, persistent heterogeneous module states, spatial correlations, bond loss, functional-cluster fragmentation, history dependence, and critical slowing. These predictions provide a mechanistic language for describing how radiation may reorganize retinal signaling beyond simple amplitude suppression.

The spin-glass and percolation layer is therefore not merely an abstract mathematical extension. It offers a biologically motivated coarse-grained representation of how injury may alter the cooperative organization of retinal populations. In this formulation, radiation can weaken, randomize, and delete effective inter-module couplings, producing a state in which local function may persist while global coordination deteriorates. This is potentially relevant to ocular radiation injury, where visual dysfunction may reflect not only loss of local response amplitude but also disruption of coordinated signaling across retinal layers and cell classes. The model consequently frames retinal injury as a possible progression from coherent collective response toward heterogeneous, frustrated, and fragmented network behavior, while recognizing that the present amplitude data do not by themselves establish that collective mechanism.

The present work also motivates specific future experiments. The most direct prospective test of the fitted injury law would combine dose escalation with network-sensitive measurements. Doses bracketing the predicted CONV threshold near 24 Gy, together with FLASH observations above 26 Gy, would test the steepness of the reduced injury law and the predicted preservation window. The retained-gain criterion should be specified before analysis and evaluated in independent eyes; a substantially different isoeffect ratio should lead to revision of the reduced injury law rather than post hoc redefinition of the threshold. In parallel, full ERG waveforms and implicit times, repeated-stimulus protocols, multifocal ERG, multielectrode-array recordings or calcium imaging, OCT, and histopathology could test the temporal, spatial, and structural predictions of the full model. Inverse-Ising or maximum-entropy analysis of simultaneous recordings could further determine whether radiation broadens the effective coupling distribution, increases frustration, or produces persistent heterogeneous states. Together, these experiments would directly assess whether the collective features predicted by the model are present in retinal responses.

Several limitations should be kept in mind. The current fits use group-mean amplitude responses, limiting sensitivity to eye-level covariance, waveform dynamics, and spatial organization. The scale-balanced $\mathcal L^*$, AIC$^*$, and BIC$^*$ are therefore screening measures rather than definitive evidence for or against a collective mechanism. Likewise, the $N=200$ units are mesoscopic functional modules, $X_i$ is a latent injury coordinate, the follow-up bridge is phenomenological, and the RC elements represent coarse-grained ion-channel conductances rather than literal anatomical circuits. The finite-size percolation scale also depends on the selected graph ensemble and bond definition, so it should be interpreted as a model-dependent quantity rather than a universal biological threshold. More broadly, the model has been calibrated to normal-retina functional data from healthy mice and cannot yet establish tumor-control equivalence, clinical retinal tolerance, or patient-specific treatment constraints.

Overall, the main contribution of this work is to provide a unified framework that connects radiation dose and dose rate, retinal injury, ERG response, and network-level disorder within a single multiscale formulation. The reduced benchmark offers a practical predictive summary of the dose-response behavior, while the full causal model provides a mechanistic hypothesis for how local injury may propagate into collective dysfunction. Together, these results support ERG-based modeling not only as a descriptive tool, but also as a platform for prospective dose selection and future network-sensitive validation in ocular radiation biology and, after appropriate experimental and translational validation, for informing studies aimed at preserving visual function during radiation treatment of the eye.

\section{Conclusion}
In this work we develop a causal $N=200$ multiscale framework connecting ocular radiation delivery to Hodgkin--Huxley-inspired, ion-channel-grounded phototransduction, a-wave dynamics, delayed post-photoreceptor signaling, and collective retinal-network organization. The two complementary fitting approaches show that the present ERG amplitude measurements strongly constrain the local injury-to-ERG response while providing more limited independent resolution of the network terms. Current ERG amplitude data support the causal radiation-injury-to-ERG framework, while additional observables are required to test and distinguish the predicted spin-glass and percolation contributions to collective retinal dysfunction. As a prospective extension, the reduced benchmark predicts a model-based functional DMF of approximately 1.21 and, for an exploratory 50\% retained-gain criterion, a conservative combined ERG threshold of approximately 28.6 Gy for FLASH versus 23.7 Gy for CONV. These values are hypotheses for experimental validation, not clinical dose constraints. Full waveform, timing, longitudinal, spatial, correlation-sensitive, structural, and dose-escalation measurements provide direct routes for testing both the predicted preservation window and the collective-network mechanisms.

%\section*{Data and Code Availability}
%The numerical implementation of the reduced and full multiscale models was performed in fixed-form Fortran 77. The computational source and model parameters can be provided with the accompanying reproducibility package.

\section*{Author Contributions}
R.A. conceived and developed the mathematical-physics framework, causal ERG formulation, radiation-injury model, spin-glass/percolation extension, and computational implementation. M.R. contributed the ocular FLASH irradiation and dosimetry context, interpretation of the Johns Hopkins experimental system, and scientific guidance regarding experimental calibration. 

\section*{Ethical Statement}
No new human or animal experiments were performed for this computational analysis. The experimental constraints derive from previously conducted ocular irradiation studies and associated data products.

\section*{Funding}
RA acknowledges support by the American Cancer Society Diversity in Cancer Research Institutional Development Grant (DICRIDG-21-074-01-DICRIDG) at Howard University.

\section*{Competing Interests}
Authors have no competing interests.

\end{document}